\documentclass[twocolumn,superscriptaddress,aps,preprintnumbers,amsmath,amssymb,footinbib,prl]{revtex4}
\usepackage{graphicx}
\usepackage{dcolumn}
\usepackage{bm}
\usepackage{hyperref}
\def\slashchar#1{\setbox0=\hbox{$#1$} 
\dimen0=\wd0 
\setbox1=\hbox{/} \dimen1=\wd1 
\ifdim\dimen0>\dimen1 
\rlap{\hbox to \dimen0{\hfil/\hfil}} 
#1 
\else 
\rlap{\hbox to \dimen1{\hfil$#1$\hfil}} 
/ 
\fi}

\def\b{\beta}

\def\d{\delta}

\def\g{\gamma}

\def\l{\lambda}

\def\s{\sigma}

\def\G{\Gamma}

\def\beq{\begin{eqnarray}}
\def\eeq{\end{eqnarray}}

\newcommand{\vev}[1]{ \left\langle {#1} \right\rangle }

\begin{document}
\title{Breit--Wigner Enhancement of Dark Matter Annihilation}

\author{Masahiro Ibe}
\affiliation{%
SLAC National Accelerator Laboratory, Menlo Park, CA 94025
}%
\author{Hitoshi Murayama}
\affiliation{Institute for the Physics and Mathematics of the
  Universe, University of Tokyo, Kashiwa 277-8568, Japan}
\affiliation{%
  Department of Physics, University of California, Berkeley, CA 94720}
\affiliation{%
  Theoretical Physics Group, Lawrence Berkeley National Laboratory,
  Berkeley, CA 94720 }
\author{T.~T.~Yanagida}
\affiliation{Department of Physics, University of Tokyo, Tokyo
  113-0033, Japan.} 
\affiliation{Institute for the Physics and Mathematics of the
  Universe, University of Tokyo, Kashiwa 277-8568, Japan}

\begin{abstract}
  We point out that annihilation of dark matter in the galactic halo
  can be enhanced relative to that in the early universe due to a
  Breit--Wigner tail, if the dark matter annihilates through a pole
  just below the threshold.  This provides a new explanation to the
  ``boost factor'' which is suggested by the recent data of the
  PAMELA, ATIC and PPB-BETS cosmic-ray experiments.
\end{abstract}

\date{\today}
\maketitle
\preprint{SLAC-PUB-13479}
\preprint{IPMU08-0097}

Dark matter of the universe has been discussed since 1933, yet its
nature still remains elusive\,\cite{Murayama:2007ek}.  Seventy-five
years later we only managed to restrict its mass between $10^{-31}$
and $10^{50}$~GeV, demonstrating our lack of understanding.  However,
the thermal relic of a Weakly Interacting Massive Particle (WIMP) is
arguably best theoretically motivated because it has the same mass
scale as the anticipated new physics that would explain why our
universe is a superconductor (electroweak symmetry breaking).  Hopes
are high to discover WIMPs at the forthcoming LHC experiments, to detect
them directly in sensitive underground experiments, as well as to
observe signals of WIMP annihilations from the galactic center or the
halo in high-energy cosmic rays.

Recent observations of the PAMELA\,\cite{Adriani:2008zr},
ATIC\,\cite{:2008zz}, and PPB-BETS\,\cite{Torii:2008xu} experiments
strongly suggest the existence of a new source of positrons (and
electrons) in cosmic rays.  The most interesting interpretation of
these results is the annihilation of the dark matter with a mass at
the TeV scale.  However, such interpretation requires that the
annihilation cross section of the dark matter in the galactic halo is
much larger (by a factor of $O(100)$) than the one appropriate to
explain the dark matter relic density precisely measured by the WMAP
experiment\,\cite{Komatsu:2008hk}.

The enhancement of the dark matter annihilation in the galactic halo
is called a ``boost factor.''  So far, there have been several
proposals to explain the origin of the boost factor both from
astrophysics such as the enhanced local dark matter density, and from
particle physics such as the Sommerfeld enhancement due to an
attractive force among the dark matter
particles\,\cite{ArkaniHamed:2008qn}.

In this letter, we propose a new explanation of the boost factor.  We
consider the dark matter which annihilates via a narrow Breit--Wigner
resonance just below the threshold.  When the resonance mass is just
below twice the dark matter mass, the annihilation cross section
becomes sensitive to the velocity of the dark matter.  In such a case,
the time evolution of the dark matter abundance is quite different
from the one in the usual non-resonant case of annihilation, and we
find that the annihilation cross section in the halo is enhanced
compared to what is inferred from the relic abundance.  As we will
show, the cross section required from the dark matter density can be
large enough to explain the PAMELA, ATIC, and PPB-BETS results, and
hence, we do not need in our proposal any additional boost factor due
to an overdense region in the halo or the Sommerfeld enhancement.


{\it Cross Section Just Above a Pole.}\/ In this study, we assume that
the dark matter with mass $m$ annihilates via a narrow resonance
$R$.  For a simplicity, we consider a scalar resonance, although
generalization to arbitrary spins is straight forward.  The general
formula for the scattering cross section via a resonance $R$ is given
by
\begin{eqnarray}
  \sigma = \frac{16\pi}{E_{\rm cm}^2\bar{\beta}_i\beta_i} 
  	\frac{M^2 \Gamma^2}{(E_{\rm cm}^2-M^2)^2 + M^2 \Gamma^2}
	B_i B_f,
\end{eqnarray}
where $M$ and $\Gamma$ are the mass and the decay rate of the
resonance $R$, respectively.  Two body initial and final states are
assumed and $\bar{\beta}_i=\sqrt{1-4m^2/M^2}$ is the initial state
phase space factor evaluated on the resonance while
$\beta_i=\sqrt{1-4m^2/E_{\rm cm}^2}$ at the center of mass energy of
the collision.  The branching fractions of the resonance into the
initial and final states are denoted by $B_{i}$ and $B_{f}$,
respectively.  Note that since we are assuming an unphysical pole,
{\it i.e.}\/, $2m>M$, $B_{i}$ and $\bar\b_{i}$ are not physical and
should be understood as analytic continuations of those quantities
from the physical region, $2m<M$.  Even so, a combination
$B_{i}/\bar\b_{i}$ is well-defined in both regions, and hence, the
above cross section is also well-defined even in the unphysical pole
case.

The dark matter annihilation in the early universe must be thermally
averaged.  On the other hand, the dark matter annihilation in the
galactic halo is averaged over the velocity distribution, which can be
approximated by the Maxwellian distribution.  In both cases, the dark
matter is non-relativistic.  Therefore, in either case, we can use the
Gaussian average,
\begin{eqnarray}
  \label{eq:gauss}
  \langle \sigma v_{\rm rel} \rangle
  = \frac{1}{(2\pi v_0^2)^3}
  \int d \vec{v}_1 d\vec{v}_2 e^{-(\vec{v}_1^2 + \vec{v}_2^2)/2v_0^2}
  2\sigma \beta_i,
\end{eqnarray}
where $\vec{v}_{1,2}$ are the velocities of the initial states and we
have used $v_{\rm rel} = 2 \b_{i}$.  The non-relativistic
approximation gives
\begin{eqnarray}
  E_{\rm cm}^2 = 4m^2 + m^2 (\vec{v}_{\rm rel})^2, \quad (\vec{v}_{\rm rel} 
  = \vec{v}_{1}-\vec{v}_{2}). 
\end{eqnarray}

\begin{figure}[t]
  \begin{center}
    \includegraphics[width=0.9\linewidth]{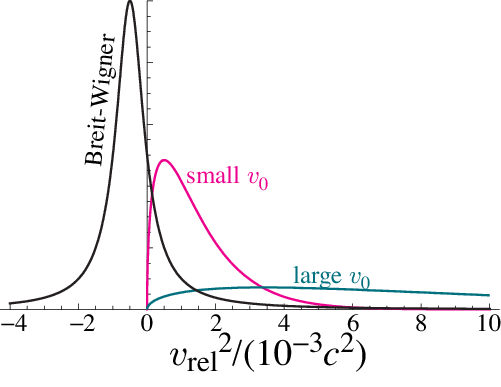}
  \end{center}
  \caption{A schematic plot that shows dispersion in relative velocity
    $v_{\rm rel}^2$ and an unphysical pole in the cross section at
    $v_{\rm rel}^2 < 0$ (below threshold).  It is clear that a smaller
    dispersion $v_0$ gives a larger overlap with the Breit--Wigner
    tail in the cross section and hence an enhanced averaged cross
    section.}
  \label{fig:schematic}
\end{figure}

Now, let us consider the annihilation process near a narrow resonance,
{\it i.e.}\/,
\begin{eqnarray}
  M^2 = 4 m^2 (1-\delta),\quad |\delta|\ll 1.
\end{eqnarray}
Note that positive $\delta$ implies that the pole is just below the
threshold of the dark matter annihilation.  With this notation, we can
rewrite the above cross section as,
\begin{eqnarray}
  \sigma = \frac{16\pi}{M^{2}\bar{\beta}_i\beta_i} 
  \frac{ \g^2}{(\d + \vec{v}_{\rm rel}^{2}/4)^2 +  \g^2}
  B_i B_f,
\end{eqnarray}
where we have defined,
\begin{eqnarray}
  \g = {\G}/{M}.
\end{eqnarray}
Furthermore, we have verified that we can approximate the Gaussian
integral reasonably well by
\begin{eqnarray}
  \label{eq:app}
  \vev{\s v_{\rm rel}}\simeq 
  \frac{32\pi}{M^{2}\bar{\beta}_i} 
  \frac{ \g^2}{(\d + \xi\,v_{0}^{2})^2 +  \g^2}B_i B_f,
\end{eqnarray}
where a parameter $\xi \approx 1/\sqrt{2}$ gives the best fit to the
numerical results for $v_{0}\ll 1$ and $\d>0$.  This expression shows
that the denominator is dominated by the $\vec{v}_{\rm rel}^{2}$ term
for $|\d|, \g \ll \xi\, v_{0}^{2}$, while the other terms dominate
when the velocity is much smaller.  Therefore, the cross section is
sensitive to the parameters $\d$ and $\g$, and have an enhanced
behavior at the lower temperature for small $\d$ and $\g$.  This
main point of this letter can be seen easily in a schematic plot in
Fig.~\ref{fig:schematic}.

Note that the approximation given in Eq.\,(\ref{eq:app}) is not a
good one for the dark matter with a rather large velocity, {\it
  i.e.}\/, $v_{0}\gtrsim 0.1$\,\cite{Griest:1990kh}.  The
approximation also becomes worse around the pole in the physical
region, possible if $\d<0$.  In the following analysis, we mainly
consider the unphysical pole $\d>0$.  As we will see below, for this
case, the resultant dark matter density is mostly determined by the
dynamics of the dark matter at $v_{0}\ll 1$, and hence the
approximated cross section works quite well.  We will later also briefly
discuss the case of a physical pole, $\d<0$.

{\it Time Evolution of Dark Matter Density.}\/ Now let us consider the
time evolution of the relic density of the dark matter.  The most
notable feature of our proposal is that the annihilation process does
not freeze out even after the ``freeze-out'' time for the usual
non-resonant annihilation models.  That is, the interaction rate can
be larger than the Hubble expansion rate even after a ``freeze-out''
time, since the cross section is enhanced at the lower temperature.
In the following, to avoid confusion, we define the ``freeze-out''
time $\tilde x_{f}$ by the usual meaning at which the yield of the
dark matter $Y=n/s$ starts deviating from the value in the thermal
equilibrium, {\it i.e.}\/, $Y-Y_{\rm EQ}=O(Y_{\rm EQ})$.  As we will
show, the actual freeze-out time is much later than $\tilde x_{f}$.
In this section, we mainly consider the unphysical pole, $\d>0$.   
 
Following Ref.~\cite{Kolb:1988aj}, section\, 5.2, we write down the
Boltzmann equation of the yield of the dark matter \footnote{The
  analysis of the Boltzmann equation near a resonance has been
  developed in Ref.~\cite{Griest:1990kh}.},
\begin{eqnarray}
  \label{eq:boltzmann}
  \frac{dY}{dx} = - \frac{\l}{x^{2}}\, 
  \frac{(\d^{2}+\g^{2})}{ (\d + \xi\, x^{-1})^{2}+\g^{2}} \,
  (Y^{2}-Y_{\rm EQ}^{2}).
\end{eqnarray}
Here, we have used the following definitions
\begin{eqnarray}
  \l &=& \left[ \frac{s(T)}{H(T)}\right]_{T = m} \vev{\s v_{\rm rel}}_{T=0}
  = \sqrt{\frac{8 \pi^{2}}{45 }g_{*}} \,M_{\rm PL}
  \,m\,\sigma_{0}\, ,
  \nonumber \\
  \sigma_{0} &=& \vev{\s v_{\rm rel}}_{T=0} = 
  \frac{32\pi B_i B_f}{M^{2}\bar{\beta}_i}
  \frac{\gamma^2}{\delta^2 + \gamma^2}\, ,
  \cr
  Y_{\rm EQ} &=& \frac{45}{4\sqrt{2}
    \, \pi^{7/2}} 
  \left(\frac{g_{i}}{g_{*}} \right)
  x^{3/2} e^{-x},\cr
  x&=& \frac{m}{T} = v_{0}^{-2}.
\end{eqnarray}
The parameter $g_{*}$ ($g_{i}$) is the number of the degrees of
freedom for massless particles (dark matter), respectively.  Note that
we have used the reduced Planck scale $M_{\rm PL} \simeq 2.4\times
10^{18}$\,GeV.
 
As we have defined, the ``freeze-out'' time $\tilde x_{f}$ is
determined by $Y-Y_{\rm EQ}=O(Y_{\rm EQ})$, and hence, the value of
$\tilde x_{f}$ is not so sensitive to the parameters $\d$ and $\g$
and mainly determined by the exponential suppression factor in $Y_{\rm
  EQ}$.  Thus, we can expect that the value of $\tilde x_{f}$ is
comparable to the freeze-out temperature $x_{f}$ in the usual
non-resonant annihilation models, {\it i.e.}\/, $\tilde x_{f}\approx
x_{f}\approx O(10)$.

Unlike the non-resonant case where the annihilation cross section
stays constant $\langle \sigma v_{\rm rel}\rangle$ once
non-relativistic, however, the annihilation cross section here increases as
temperature drops.
As a result, the annihilation process does not freeze out even for
$x>\tilde x_{f}$ and the dark matter keeps annihilating until the
temperature comes down to
\begin{eqnarray}
  x_{b} \simeq \xi^{-1}\times \max\left[\d, \g \right]^{-1} \gg \tilde x_{f}.
\end{eqnarray}
Below this temperature, Eq.~(\ref{eq:boltzmann}) reduces to
\begin{eqnarray}
  \frac{dY}{dx} = - \frac{\l}{x^{2}} \,Y^{2},
\end{eqnarray}
and we obtain an asymptotic solution,
\begin{eqnarray}
  \label{eq:sol}
  Y _{ \infty} \simeq \frac{1}{\l} \,x_{b} \simeq \frac{1}{\l} \times
  \max\left[\d, \g \right]^{-1}. 
\end{eqnarray}

\begin{figure}[t]
  \begin{center}
    \includegraphics[width=0.9\linewidth]{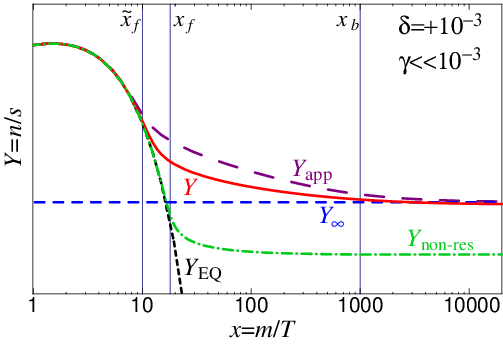}
  \end{center}
  \caption{The time evolution of the yield $Y$ of the dark matter in
    terms of the parameter $x = m/T$ for given values of $\d$ and
    $\g$ (the solid line).  In this figure, we assume $\d>0$ and
    the pole is not in the physical region.  The long-dashed line
    labeled $Y_{\rm app}$ is the evolution with the approximated cross
    section in Eq.\,(\ref{eq:app}).  The dashed line labeled
    $Y_\infty$ is the asymptotic solution $Y_{\infty}$ given in
    Eq.\,(\ref{eq:sol}).  The short-dashed line represents the
    equilibrium yield $Y_{\rm EQ}$.  The dash-dotted line labeled
    $Y_{\rm non-res}$ shows the time evolution of the yield in the
    usual (non-resonant) freeze-out assuming the same cross section at
    the low temperature $\sigma_{0}$ used in $Y$ (see
    Eq.\,(\ref{eq:sol2})).  The boost factor is the asymptotic value
    of $Y/Y_{\rm non-res}$.}
  \label{fig:boltzmann}
\end{figure}

In the Fig.\,\ref{fig:boltzmann}, we show the time evolution of the
yield $Y$ of the dark matter for a given parameter set.  Here, we have
used the numerical result of the Gaussian average of the cross section
\footnote{In the computation, we have used $\l = 10^{9}$.  The
  qualitative behavior of the time evolution does not depend on $\l$,
  although the values of $x_{f}$ and $\tilde x_{f}$ slightly depend
  on $\l$.  }.  We also show the time evolution of the yield with the
approximate cross section given in Eq.\,(\ref{eq:app}), $Y_{\rm app}$.
As we have expected, the yield deviates from $Y_{\rm EQ}$ at $\tilde
x_{f}=O(10)$, while the actual freeze-out occurs at $x_{b} \gg \tilde
x_{f}$.  From the figure, we see that $Y<Y_{\rm app}$ during $\tilde
x_{f}<x<x_{b}$.  This means that the averaged cross section at $\tilde
x_{f}$ is somewhat larger than the approximate one, while the final
result is determined by the late time dynamics where the approximate
cross section works well.  The figure shows that the above approximate
asymptotic solution gives a good estimate of the resultant yield of
the dark matter.

It is worth comparing the asymptotic solution in Eq.\,(\ref{eq:sol})
with the asymptotic solution in the usual non-resonant ($S$-wave)
annihilation models\,\cite{Kolb:1988aj},
\begin{eqnarray}
  \label{eq:sol2}
  Y_{\infty}\simeq \frac{1}{\l}\, x_{f},
\end{eqnarray}
with $x_{f} = O(10)$.  These two solutions show that the dark matter
abundance in our proposal is larger by a factor
\begin{eqnarray}
  \frac{x_{b}}{x_{f}} \simeq 
  \frac{\max\left[ \d, \g \right]^{-1}}{O(10)}
\end{eqnarray}
when we assume the same cross section at the zero temperature for both
models.  In the Fig.\,\ref{fig:boltzmann}, we have showed the time
evolution of the yield in the usual non-resonant annihilation models
assuming the same cross section at the zero temperature $\sigma_{0}$
used in $Y$ ({\it i.e.}\/, $\l = 10^{9}$).  As expected, the yield in
the non-resonant annihilation is more suppressed compared to the
resonant annihilation.  The physical reasons of this enhancement are
as follows.  First, in our proposal, the cross section is suppressed
by $(\tilde x_{f}/x_{b})^{2}$ at $\tilde x_{f}$, which results in a
relatively larger abundance in a period of $x>\tilde x_{f}$.  Second,
the annihilation process is relatively less effective during $ \tilde
x_{f}<x<x_{b}$ compared to the usual annihilation case,
although the annihilation does not freeze out in that period.\\

Before closing this section, we mention the model with a physical pole,
{\it i.e.}\/, $\d<0$.  In this case, the cross section given in
Eq.\,(\ref{eq:app}) poorly approximates the thermal averaged cross
section\,\cite{Griest:1990kh}.  Especially, the thermal average can
pick up the cross section at the pole, $v_{\rm rel}^{2}= 4 |\delta|$ when the
temperature is rather high, {\it i.e.}\/, $x^{-1}\gg |\delta|$. 
Thus, the averaged cross section can be much higher than that expected in the
unphysical pole where the cross section is suppressed by
$x^{2}(\d^{2}+\g^{2})$ at high temperature.  
Thus, the annihilation cross section
at $\tilde x_{f}$, which typically satisfies $\tilde x_{f}^{-1}\gg
|\d|,\g$, is much larger than that in the case of the unphysical pole. 
Therefore, the yield enhancement seen in the unphysical pole is 
much smaller in this case.


\begin{figure}[t]
  \begin{center}
    \includegraphics[width=0.9\linewidth]{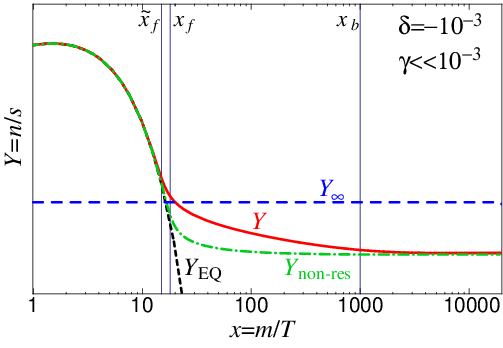}
  \end{center}
  \caption{The time evolution of the yield of the dark matter $Y$ for
    $\d<0$ (the solid line).  Everything else is the same as in
    Fig.~\ref{fig:boltzmann}.  There is practically no boost for this
    parameter set.  }
  \label{fig:boltzmann2}
\end{figure}

{\it Effective Boost Factor.}\/ Finally, let us work out the boost
factor for annihilation in the galactic halo.  As we have seen, the
dark matter abundance is enhanced by a factor of $(x_{b}/x_{f})$ for
$\d>0$ compared to the usual non-resonant annihilation models for a
given annihilation cross section at the zero temperature,
$\sigma_{0}$.  This means that the cross section $\sigma_{0}$ must be
larger than the one expected in the usual models to reproduce the
observed dark matter density.

More explicitly, the yield given in Eq.\,(\ref{eq:sol}) is translated
to the mass density parameter,
\begin{eqnarray}
  \Omega_{\rm DM} h^{2} 
  \simeq 0.1 \times \left(\frac{10^{-9} \,{\rm GeV}^{-2}}{\sigma_{0}}
  \right) \left(\frac{x_{b}}{x_{f}}\right), 
\end{eqnarray}
where we have used $x_{f}\simeq 20$ and $g_{*}\simeq 200$.  Therefore,
the observed dark matter density $\Omega h^{2}\simeq 0.1$ requires
\begin{eqnarray}
  \label{eq:CS}
  \sigma_{0}\simeq 10^{-9}\,{\rm GeV}^{-2} \times
  \left(\frac{x_{b}}{x_{f}}\right), 
\end{eqnarray}
which is much larger than what is expected in the usual annihilation
case, {\it i.e.}\/,
\begin{eqnarray}
  \label{eq:CS2}
  \sigma_{0} \simeq 10^{-9}\,{\rm GeV}^{-2}. 
\end{eqnarray}

In the galactic halo, the average velocity is given by $v_{0}\simeq
10^{-3}$, and the cross section is well approximated by the one at the
zero temperature as long as $v_{0}^{2}\ll \d,\g$.  Thus, we can
achieve the large annihilation cross section suggested by the PAMELA,
PPB-BETS, and ATIC anomalies, $\vev {\s v_{\rm rel}}_{T=0} =
O(10^{-(6-7)})$\,GeV$^{-2}$ for $\d,\g\lesssim 10^{-3}$.
Therefore, in our proposal, we can explain the large annihilation
cross section in the galactic halo without other boost factors.

\begin{figure}[tbh]
  \begin{center}
    \includegraphics[width=0.8\linewidth]{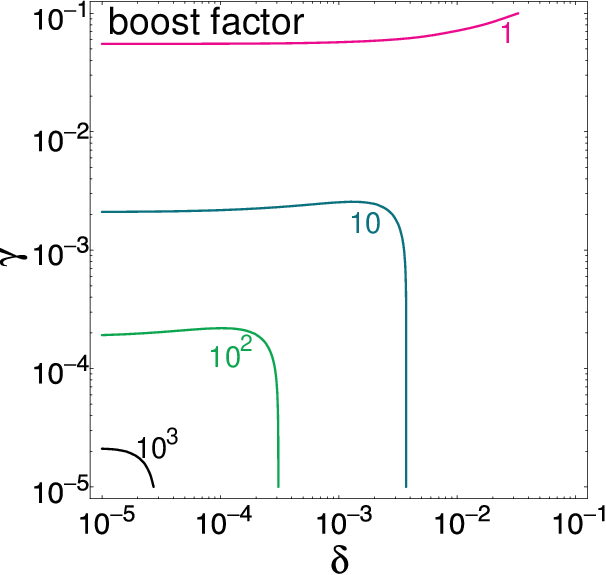}
  \end{center}
  \caption{The boost factor on the $(\delta,\gamma)$ plane.  Thermal
    average is done numerically without relying on the approximation
    Eq.~(\ref{eq:app}). }
  \label{fig:BFP2}
\end{figure}

For convenience, we could define an effective boost factor as the
ratio between the cross sections in the usual and our models.  From
Eqs.\,(\ref{eq:CS}) and (\ref{eq:CS2}) we obtain
\begin{eqnarray}
  {\rm BF} \simeq \frac{x_{b}}{x_{f}} \simeq\frac{\max\left[ \d,
      \gamma  \right]^{-1} } {x_{f}} \simeq 
  \frac{\max\left[\d, \gamma  \right]^{-1} }{O(10)}.
\end{eqnarray}
In this way, we can explain the boost factor of $O(100)$, in a model
with $\d,\g = O(10^{-3})$, as seen in Fig.~\ref{fig:BFP2}.

{\it Discussion.}\/
In this letter, we showed that the boost factor required in recent
observations of cosmic ray electrons and positrons can be obtained if
the dark matter annihilates via a narrow resonance just below the
threshold.  Because the annihilation in the early universe is
suppressed by the Breit--Wigner tail, the observed dark matter density
requires a larger-than-normal cross section which can be consistent
with the PAMELA, ATIC and PPB-BETS results.  The required cross
section is achieved for $4m^2-M^2, M \Gamma \lesssim 10^{-3} M^2$.

Note that the small decay width $\Gamma \ll 10^{-3} M$ can be achieved
rather easily, when the resonance $R$ is a weakly coupled particle.
In some models, a coincidence in masses can also be naturally
realized.  For example, if the dark matter is the lightest
Kaluza-Klein (KK) particle and the resonance is at the second KK
level, we have the relation $M=2m$ at the
tree-level\,\cite{Kakizaki:2005en} \footnote{In the explicit model
  given in the reference, the resonance mainly decays into a top pair
  which is not favored by the null results of the excess of the
  anti-proton flux\,\cite{Adriani:2008zr,:2008zz,Torii:2008xu}.  }
(see also a recent discussion in Ref.~\cite{:2008zz}).  Thus, in such
models, we may well have a small mass splitting $\delta$ as a result
of small radiative corrections to the tree-level mass relation.

\acknowledgements

The work of M.I. was supported by the U.S. Department of Energy under
contract number DE-AC02-76SF00515.  The work of H.M. and T.T.Y. was
supported in part by World Premier International Research Center
Initiative (WPI Initiative), MEXT, Japan.  The work of H.M. was also
supported in part by the U.S. DOE under Contract DE-AC03-76SF00098,
and in part by the NSF under grant PHY-04-57315.

\end{document}